\documentclass[aps,amsmath,amssymb,superscriptaddress,twocolumn,longbibliography]{revtex4-1}

\usepackage{amsmath}
\usepackage{graphicx}
\usepackage[dvipsnames]{color}
\usepackage{hyperref}
\usepackage{psfrag}
\usepackage{stmaryrd}
\usepackage{amssymb}
\usepackage{wasysym}
\usepackage{float}
\usepackage{color}

\usepackage[T1]{fontenc}
\usepackage[french,english]{babel}
\usepackage{lmodern}
\usepackage{graphicx} 
\usepackage{times}
\usepackage{amsmath,amssymb}
\usepackage{color}

\newcommand{\bnabla}{\boldsymbol{\nabla}}

\newcommand{\la}{\left<}
\newcommand{\ra}{\right>}

\begin{document}

\title{Two-layer baroclinic turbulence with arbitrary layer depths}

\author{Gabriel Hadjerci}
\email[]{gabriel.hadjerci@cea.fr}
\affiliation{Universit\'e Paris-Saclay, CNRS, CEA, Service de Physique de l'Etat Condens\'e, 91191 Gif-sur-Yvette, France.}
\author{Basile Gallet}
\affiliation{Universit\'e Paris-Saclay, CNRS, CEA, Service de Physique de l'Etat Condens\'e, 91191 Gif-sur-Yvette, France.}

\begin{abstract}

While heat transport by baroclinic turbulence in oceans and planetary atmospheres is well described by a two-layer model, the relative depth of the two layers varies greatly depending on the situation of interest, making it an important parameter governing the transport properties of the system. 
Focusing on the low-drag turbulent regime, we extend the vortex-gas scaling theory to address the case of arbitrary layer depths. To wit, we map the arbitrary-layer-depth system onto an equivalent equal-depth system with rescaled parameters, establishing the asymptotic validity of the mapping for weak bottom drag. This approach leads to quantitative predictions for the turbulent transport by two-layer baroclinic turbulence with arbitrary layer depths, without additional free parameters. We validate these predictions using an extended suite of numerical simulations with either linear or quadratic bottom drag.
\end{abstract}

\maketitle

\textit{Introduction.--} Oceans and planetary atmospheres host large-scale meridional temperature gradients that induce turbulence through the baroclinic instability mechanism. The resulting turbulent flow enhances heat transport across latitudes, thus mitigating the temperature difference between the equator and the poles. In the Earth atmosphere the typical scale of the eddies -- the Rossby deformation radius -- is large and well-resolved by global climate models. By contrast, ocean mesoscale eddies have a core diameter of tens of kilometers. {Ocean eddies thus} remain unresolved in most global climate models and the associated transport needs to be parametrized \cite{held_macroturbulence_1999,venaille_baroclinic_2011,hewitt_resolving_2020,yankovsky_influences_2022}. {In particular, eddy-induced turbulent transport in the Southern Ocean is one of the key processes contributing to the stratification of the various ocean basins and to the slow meridional overturning circulation~\cite{Wolfe2010,Nikurashin11,Nikurashin12}.}

{While the scale separation between the ocean eddies and the larger ocean basins poses a challenge for direct resolution by global climate models, this same scale separation is the starting point of a useful asymptotic approach for physicists~\cite{Pavan}. Indeed, a flow with a small turbulent mean-free path (also known as the `mixing-length') induces transport of diffusive nature. This motivates the study of horizontally periodic `patches' of ocean, from which one infers the turbulent eddy diffusivity to be included in larger-scale ocean models (see Ref.~\cite{gallet_vortex_2020} for a pedagogical illustration). The physicist's approach further} consists in reducing the problem to its bare minimum \cite{lucarini_mathematical_2014}. This led Phillips to introduce in 1954 the two-layer quasi-geostrophic (2LQG) model on an $f$-plane \cite{phillips_energy_1954,salmon_lectures_1998,pedlosky_geophysical_2003,vallis_atmospheric_2017}. Even for this strongly idealized configuration, however, the parameterization of the turbulent transport proves surprisingly challenging. Early attempts were framed in spectral space, based on the idea that energy input by baroclinic instability at a scale comparable to the deformation radius is transferred to larger scales following a Kolmogorov inverse energy cascade \cite{larichev_eddy_1995, held_scaling_1996, held_macroturbulence_1999}. While qualitatively insightful, the resulting spectral theory lacks quantitative agreement with numerical simulations~\cite{held_scaling_1996}.
The spectral-space description was further challenged by Thompson \& Young \cite{thompson_scaling_2006} who suggested that the flow is better described in physical space because of the emergence of intense coherent vortices. Gallet \& Ferrari \cite{gallet_vortex_2020} recently embraced this idea to derive a scaling theory in physical space, based on the idea that the flow can be described as a dilute gas of idealized vortices. The resulting `vortex-gas' theory provides quantitative predictions that are validated by numerical simulations of the 2LQG model with equal layer depths \cite{gallet_vortex_2020,gallet_quantitative_2021} and carry over to a 3D model with uniform density stratification \cite{gallet_transport_2022}. In parallel with the development of the vortex-gas theory, a current line of research consists in understanding how the standard cascade arguments must be corrected for if one insists on describing the flow in spectral space \cite{chang_control_2019,chang_parameter_2021,chen_revisiting_2023}. Ultimately, the theory that best describes the system (spectral- vs. physical-space approach) is arguably the one that includes the smallest number of adjustable parameters, while describing the largest number of model configurations of interest. This is the first motivation for the present study. 

{The second motivation is that any practical situation of interest departs from a uniform stratification, so that the associated two-layer model must include unequal depths for the two layers. We have not hidden our predominant interest in mesoscale ocean turbulence, which typically corresponds to an upper layer five times shallower than the lower layer~\cite{salmon_two-layer_1978,fu_nonlinear_1980}. By contrast, 
the atmosphere corresponds to an upper layer twice thicker than the lower layer \cite{fandry_two-layer_1984}.} Arbitrary layer depths complicate the problem at both the quantitative and the qualitative levels. New terms arise in the equations, inducing lower levels of barotropization \cite{flierl_models_1978,fu_nonlinear_1980,hua_numerical_1986,smith_scales_2001}, a term that refers to the tendency for the system to form an intense vertically-invariant flow~\cite{charney_geostrophic_1971,salmon_two-layer_1978}. Predicting the overall transport induced by baroclinic turbulence in a two-layer model with arbitrary layer depths is thus both a challenge to test the robustness of the relatively recent vortex-gas theory, and the natural next step to extend it to realistic situations of interest.

\begin{figure}[h!]
    \centering
    \includegraphics[width=8.5cm]{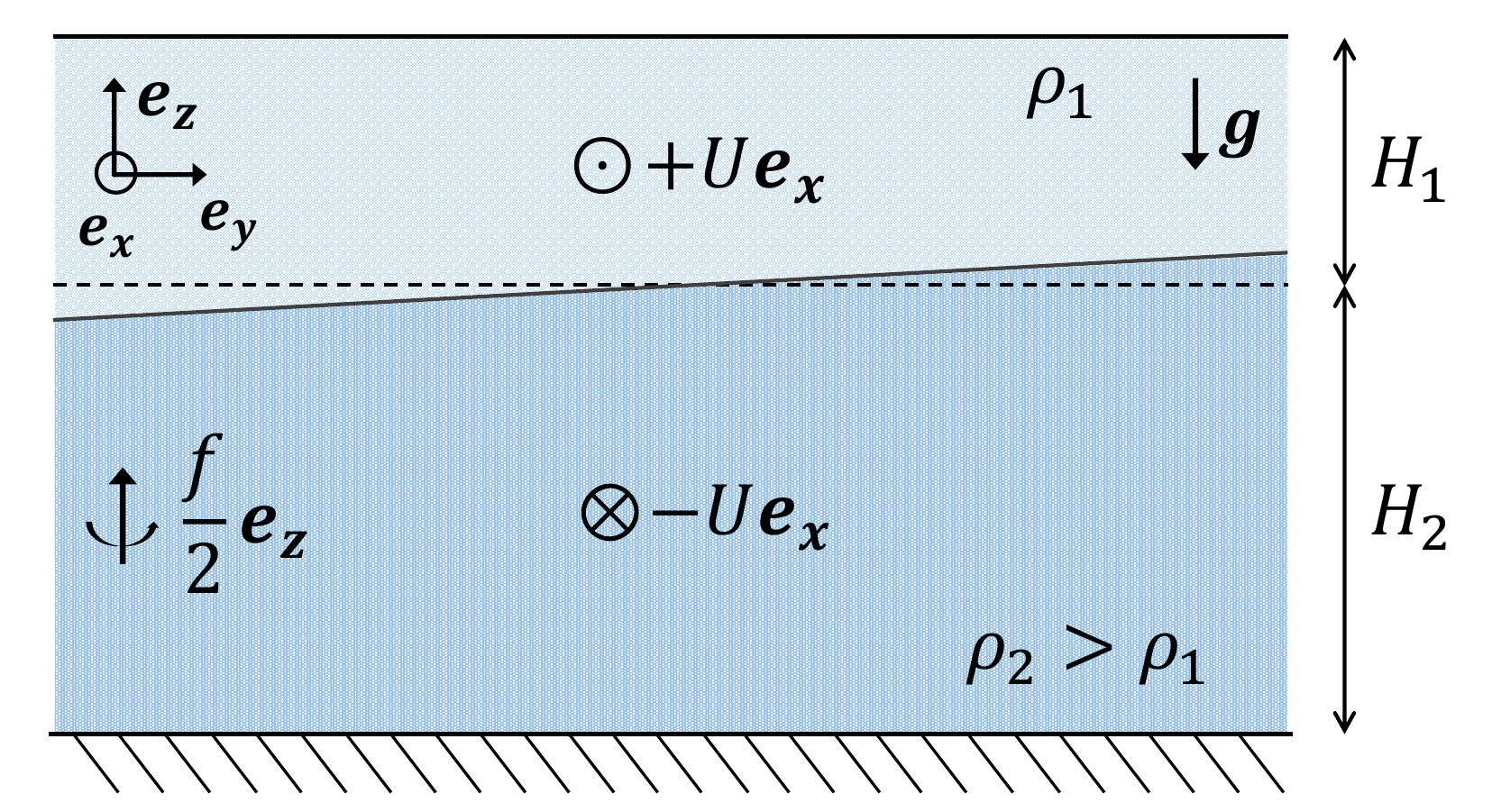}
    \caption{Two layers of fluid with densities $\rho_1$ and $\rho_2$ sit on top of one another, in a frame rotating at a rate $f/2$ around the vertical axis. The base state consists of a vertically sheared zonal flow. The interface between the layers is tilted as a consequence of thermal wind balance.}
    \label{fig:2LQG_sketch}
\end{figure}

The 2LQG system has been extensively described in the literature \cite{flierl_models_1978,salmon_lectures_1998,arbic_baroclinically_2004}. Two shallow layers of fluid sit on top of one another in a frame rotating at a rate $f/2$ around the vertical axis ${\bf e}_z$. We denote with a subscript $1$ (resp. $2$) quantities in the upper (resp. lower) layer. When the system is at rest, the total fluid depth is $H$, with $H_1$ the depth of the upper layer, $H_2=H-H_1$ the depth of the lower layer and $\alpha=H_1/H$ the relative depth of the upper layer. The small difference between the upper-layer density $\rho_1$ and the lower-layer density $\rho_2$ is associated with a reduced gravity $g'=g (\rho_2-\rho_1)/\rho_1$, where $g$ denotes the standard gravity, and we define an overall Rossby deformation radius as $\lambda=\sqrt{g' H}/(2 f)$. 
The base state consists of uniform arbitrary velocities ${\bf U}_{1;2}$ in each layer. We denote as $x$ the direction of ${\bf U}_1-{\bf U}_2$ and we introduce the shearing velocity ${U}{\bf e}_x=({\bf U}_1-{\bf U}_2)/2$. Leveraging Galilean invariance, the governing equations are then conveniently written in the frame moving at the velocity $({\bf U}_1+{\bf U}_2)/2$, which we adopt in the following. The system is represented schematically in this frame of reference in Figure~\ref{fig:2LQG_sketch}: the base velocities reduce to $+{U}{\bf e}_x$ in the upper layer and $-{U}{\bf e}_x$ in the lower layer. {Through thermal-wind balance, the shear flow is associated with a tilt of the interface: there is a meridional gradient in vertically averaged density, which is the potential energy reservoir feeding baroclinic instability.}
We denote as $\psi_{1;2}(x,y,t)$ the departure streamfunctions with respect to this base state{, the velocity fields in the two layers being ${\bf u}_1=+{U}{\bf e}_x-\bnabla \times (\psi_1 {\bf e}_z)$ and ${\bf u}_2=-{U}{\bf e}_x-\bnabla \times (\psi_2 {\bf e}_z)$.}

The potential vorticity (PV) in each layer reads $Q_{1;2}=G_{1;2} \, y + q_{1;2}(x,y,t)$, where $G_{1;2}$ are the background meridional PV gradients associated with the base state and $q_{1;2}$ are the departure PVs:
\begin{eqnarray}
G_1=\frac{{U}}{2 \alpha \lambda^2} \, & , & \quad q_1=\nabla^2 \psi_1 + \frac{\psi_2-\psi_1}{4 \alpha \lambda^2} \, ,\\
G_2=- \frac{{U}}{2 (1-\alpha) \lambda^2} \, & , & \quad q_2=\nabla^2 \psi_2 + \frac{\psi_1-\psi_2}{4 (1-\alpha) \lambda^2} \, .
\end{eqnarray}
The evolution of the system is governed by the material conservation of PV in each layer:
\begin{eqnarray}
\partial_t q_1 + {U}\partial_x q_1 +J(\psi_1,q_1) + \frac{{U}}{2 \alpha \lambda^2} \partial_x \psi_1& = & 0 \, , \label{eq:q1} \\ 
\partial_t q_2 - {U}\partial_x q_2  +J(\psi_2,q_2) - \frac{{U}}{2 (1-\alpha) \lambda^2} \partial_x \psi_2 & = & \text{drag} \, , \quad \label{eq:q2}
\end{eqnarray}
where $J(\psi,\varphi)=\partial_x(\psi)\partial_y(\varphi)-\partial_y(\psi)\partial_x(\varphi)$.

We have included a bottom drag term in the lower layer {to damp kinetic energy. Frictional dissipation in the lower layer balances the release of energy by baroclinic instability} in the equilibrated state. We consider either linear bottom drag with a coefficient $\kappa$ or quadratic drag with a coefficient $\mu$, that is $\text{drag} = - 2 \kappa {\nabla}^2 \psi_2$ or $\text{drag} =- \mu \left[ \partial_x(|\boldsymbol{\nabla} \psi_2| \partial_x \psi_2) + \partial_y(|\boldsymbol{\nabla} \psi_2| \partial_y \psi_2)  \right]$. We are interested in the statistically-steady solutions of (\ref{eq:q1}-\ref{eq:q2}) with periodic boundary conditions {in both $x$ and $y$}, in a large-enough domain for the transport properties of the flow to be independent of the horizontal extent of the domain. {Additionally, when numerically simulating the evolution equation for $q_{1;2}$ we include a small hyperdiffusive term $\nu\Delta^4 q_{1;2}$ (respectively)  to damp the small-scale vorticity filaments, making sure that the transport properties of the flow are independent of the small hyperdiffusion coefficient.}

The key quantity of interest is the effective diffusivity of the equilibrated flow {\cite{held_macroturbulence_1999,Pavan,gallet_vortex_2020}}. Denoting with angular brackets a time and horizontal area average over the domain, the PV diffusivity in each layer is obtained by dividing the eddy-induced meridional PV flux $\la q_{1;2} \, \partial_x \psi_{1;2} \ra$ by minus the background PV gradient $-G_{1;2}$. Conveniently, {after a few integrations by parts using the periodic boundary conditions,} this operation results in the same expression for the PV diffusivity in the two layers. That is, the transport of PV in both layers (as well as the heat transport, see below) is governed by a single eddy diffusivity coefficient \cite{thompson_scaling_2006,chang_control_2019,gallet_vortex_2020}:
\begin{eqnarray}
D & = & \frac{\la \psi_1 \, \partial_x \psi_2 \ra}{2 \, {U}} \, . \label{eq:defD}
\end{eqnarray}

The goal of the present study is to determine $D$ in terms of the control parameters of the problem: ${U}$, $\lambda$, $\alpha$ and the drag coefficient $\kappa$ or $\mu$. Non-dimensionalizing space and time using $\lambda$ and ${U}$, we seek the expression of the dimensionless diffusivity ${D}_*=D/({U} \lambda)$ in terms of the relative depth of the layers $\alpha$ and the dimensionless drag coefficient ${\kappa}_*=\kappa \lambda / {U}$ or ${\mu}_*=\mu \lambda$. That is, we seek the dimensionless functions ${\cal D}_l(\alpha, \kappa_*)$ and ${\cal D}_q(\alpha, \mu_*)$ such that:
\begin{eqnarray}
D_* & = & {\cal D}_l(\alpha, \kappa_*) \, \quad \text{or} \quad D_*  =  {\cal D}_q(\alpha, \mu_*) \, ,
\end{eqnarray}
for linear and quadratic drag, respectively.

\begin{figure}[h!]
    \centering
    \includegraphics[width=8.6cm]{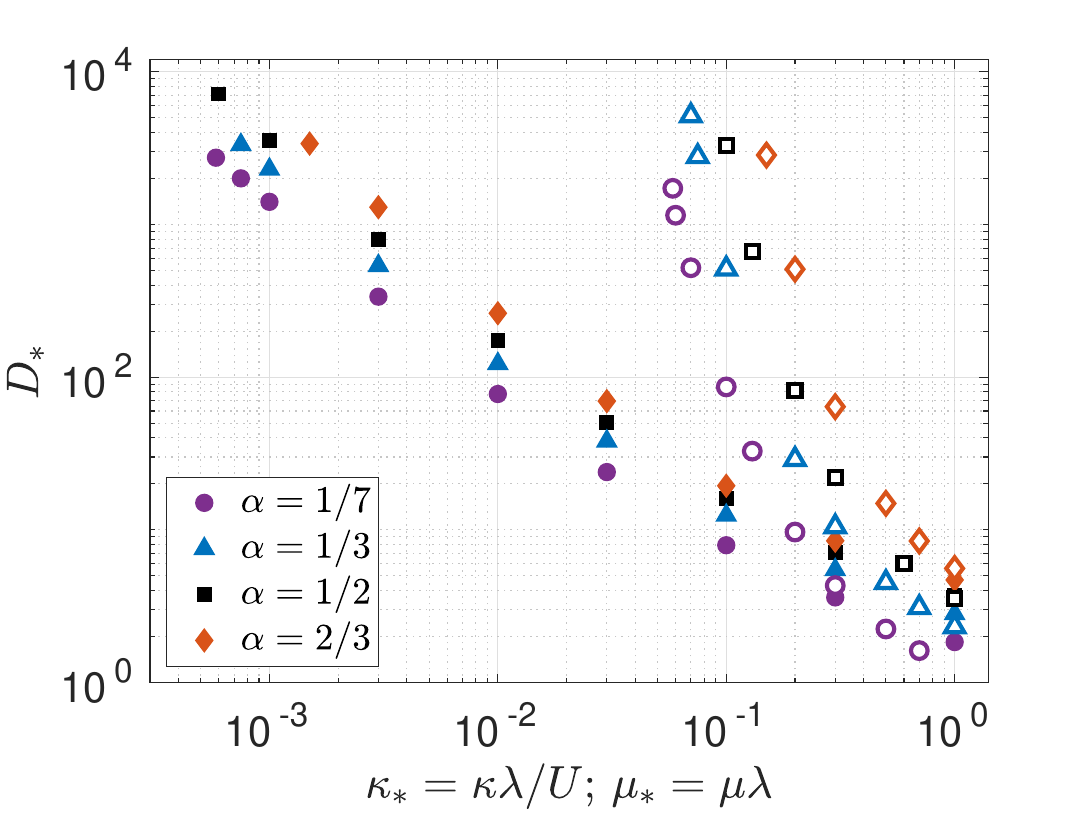}\\
    \caption{Dimensionless diffusivity as a function of the dimensionless drag coefficient for four different values of the relative layer depths, using either linear drag (open symbols) or quadratic drag (filled symbols).}
    \label{fig:raw_data}
\end{figure}

To wit, we have performed numerical simulations of equations (\ref{eq:q1}-\ref{eq:q2}) using a pseudo-spectral solver, with large enough domain size and small enough hyperviscosity for these two parameters to be irrelevant. The suite of simulations consists of sweeps of the dimensionless friction coefficient for several values of the relative layer depth $\alpha$. Once the system has reached a statistically steady state we extract the eddy diffusivity by performing the spatial and time average arising in equation~(\ref{eq:defD}). The resulting values of ${D}_*$ are shown in Figure \ref{fig:raw_data}. A striking feature is that the diffusivity varies significantly with the relative depth $\alpha$ of the layers, for otherwise constant parameters. 

\textit{Barotropic-baroclinic decomposition.--}The dynamics are better characterized by recasting the governing equations in terms of the `barotropic' streamfunction $\psi$ associated with the vertically averaged flow, $\psi=\alpha \psi_1 + (1-\alpha) \psi_2$, together with the `baroclinic' streamfunction $\tau=\sqrt{\alpha(1-\alpha)}(\psi_1-\psi_2)$:
\begin{eqnarray}
& &  \partial_t(\nabla^2 \psi) + J(\psi,\nabla^2 \psi) + J(\tau,\nabla^2 \tau) + \chi {U} \partial_x (\nabla^2 \tau)     \label{eq:fullpsi}\\
\nonumber  & & \quad \quad  \quad \quad   \quad \quad  \quad \quad   \quad \quad  \quad \quad  =   (1-\alpha) \, \text{drag} \, , \\
& &  \partial_t \left(\nabla^2 \tau - \frac{\tau}{\chi^2 \lambda^2} \right) + J\left(\psi, \nabla^2 \tau - \frac{\tau}{\chi^2 \lambda^2}  \right)   + J(\tau,\nabla^2 \psi) \quad \,  \label{eq:fullttau}  \\
\nonumber & &  \quad \quad   \quad \quad  \quad \quad  + \chi {U} \partial_x \left(\nabla^2 \psi +  \frac{\psi}{\chi^2 \lambda^2} \right) + {\cal T} =  - \frac{\chi}{2} \, \text{drag}\, ,
\end{eqnarray}
where ${\cal T}=\frac{2(1-2 \alpha)}{\chi} \left[ J(\tau, \nabla^2 \tau) + \chi {U} \partial_x(\nabla^2 \tau) \right]$ and we have introduced the short-hand notation $\chi=2\sqrt{\alpha(1-\alpha)}$.
The baroclinic streamfunction $\tau$ is sometimes referred to as the temperature variable, whose dynamics is coupled to that of the vertically averaged (or barotropic) flow. Because of thermal-wind balance, the background shear flow is associated with a background meridional gradient in vertically averaged density, visible as a tilt of the interface in Figure~\ref{fig:2LQG_sketch}. That is, there is a background meridional gradient $G_\tau=-\chi U$ for the temperature variable, and a straightforward calculation shows that the meridional heat transport by the vertically averaged flow is characterized by the same diffusivity $-\la \psi_x \tau \ra/G_\tau=D$ as PV transport.

The equal-depth case is recovered by substituting $\alpha=1/2$ (and therefore $\chi=1$) in the equations above, which also leads to ${\cal T}=0$. An important qualitative difference between the equal-depth case and the unequal-depth case is therefore the presence of the additional term ${\cal T}$ in the latter situation. To make progress on the unequal-depth case, we first briefly recall the low-drag behavior of the equal-depth system.


\textit{Vortex-gas scaling theory.--}The low-drag scaling behavior of the equal-depth system is accurately described by the vortex-gas scaling theory \cite{gallet_vortex_2020,gallet_quantitative_2021,gallet_transport_2022,hadjerci_vortex_2023} . At low drag the flow barotropizes, with most of the kinetic energy in the barotropic flow ${\bf u}=-{\bnabla}\times (\psi {\bf e}_z)$. The corresponding barotropic vorticity field resembles a dilute gas of  coherent vortices \cite{thompson_scaling_2006}, which we model as isolated vorticity patches of typical radius $r_\text{core}$ separated by a typical inter-vortex distance $\ell_{iv} \gg r_\text{core}$. The vortices have circulation $\pm \Gamma$ and they wander around as a result of mutual induction with a typical core velocity $V\sim \Gamma/\ell_{iv}$. One further assumes that the transport properties of the gas are correctly inferred by considering an isolated dipole of opposite-sign vortices. This leads to fluctuations of the baroclinic streamfunction of order $\tau \sim {U} \ell_{iv}$ and a typical diffusivity that scales as $D \sim \ell_{iv} V$. The vortex core radius was initially assumed to be of the order of the deformation radius, a reasonable assumption for moderately low bottom drag \cite{gallet_vortex_2020}. However, investigation of the very-low drag regime indicates that $r_\text{core}$ becomes significantly greater than $\lambda$ and scales as $r_\text{core} \sim \sqrt{\lambda \ell_{iv}}$ instead, as inferred from a potential vorticity conservation argument and detailed in Ref.~\cite{hadjerci_vortex_2023}. One completes the theory using two energetic arguments: first, the `slantwise-free-fall' argument consists in considering a fluid column initially at rest that travels `freely' over a typical mean free path $\ell_{iv}$. Equating the drop in potential energy of the fluid column with its final kinetic energy leads to $V/{U} \sim \ell_{iv}/\lambda$. The second energetic argument is the energy balance equation:  in stationary state, the rate of release of potential energy by baroclinic instability $D {U}^2/\lambda^2$ equals the frictional energy dissipation rate. Because of the strong barotropization, we estimate the latter using the frictional dissipation acting on the barotropic flow only, of order $\kappa \la {\bf u}^2 \ra$ for linear drag and $\mu \la {\bf u}^3 \ra$ for quadratic drag (in the equal-depth case). As detailed in Ref. \cite{hadjerci_vortex_2023}, combining these scaling relations leads to the following expressions for the dimensionless diffusivity of the equal-depth system ($\alpha=1/2$), for linear and quadratic drag respectively:
\begin{eqnarray}
{\cal D}_l(1/2, {\kappa}_*) & = & c_1 \exp \left( \frac{c_2}{{\kappa}_*}  \right) \, , \label{eq:VGlin}\\
{\cal D}_q(1/2, {\mu}_*) & = & \frac{c_3}{{\mu}_*^{4/3}} \, .\label{eq:VGquad}
\end{eqnarray}


\textit{Mapping onto an equal-depth system.--}The characterization of the equal-depth system allows us to understand the unequal-depth one in the following way. Let us first assume that the term ${\cal T}$ in equation (\ref{eq:fullttau}) is negligible in the low-drag regime, an assumption that we will justify \textit{a posteriori}. The key observation is then that, in the absence of the term ${\cal T}$, the left-hand sides of equations (\ref{eq:fullpsi}-\ref{eq:fullttau}) are exactly those of the equal-depth system, with the rescaled parameter values $\hat{U}=\chi U$ and $\hat{\lambda}=\chi \lambda$ in place of $U$ and $\lambda$, respectively. We thus expect the system to be governed by the vortex-gas scaling theory with these rescaled parameter values. In line with the vortex-gas theory, the rescaled drag coefficient is obtained by neglecting the drag force in the $\tau$ equation (\ref{eq:fullttau}) and neglecting $\tau$ in the expression of the drag force that arises in the $\psi$ equation (\ref{eq:fullpsi}). In other words, only the drag force acting on the dominant barotropic mode comes into play in the low-drag regime. The latter drag force is $- 2 \kappa (1-\alpha) {\nabla}^2 \psi$ for linear drag and $- \mu (1-\alpha)  \left[ \partial_x(|\boldsymbol{\nabla} \psi| \partial_x \psi) + \partial_y(|\boldsymbol{\nabla} \psi| \partial_y \psi)  \right]$ for quadratic drag. Once again, these terms are identical to those arising in an equal-depth system, up to a rescaling of the friction coefficient as $\hat{\kappa}=2 (1-\alpha) \kappa$ or $\hat{\mu}=2 (1-\alpha) \mu$. To summarize, provided the term ${\cal T}$ can be neglected, the unequal-depth system can be mapped onto an equal depth system with rescaled parameter values $\hat{U}$, $\hat{\lambda}$ and $\hat{\kappa}$ or $\hat{\mu}$. We conclude that the scaling-laws of the equal-depth system govern the scaling behavior of the unequal-depth system provided one uses the rescaled parameter values. That is, the temperature diffusivity $D=\la \psi_x \tau \ra / \hat{U}$ satisfies:
\begin{eqnarray}
\frac{D}{\hat{U} \hat{\lambda}} & = & {\cal D}_l\left(\frac{1}{2}, \frac{\hat{\kappa} \hat{\lambda}}{\hat{U}}\right) \, \quad \text{or} \quad  \frac{D}{\hat{U} \hat{\lambda}}  =  {\cal D}_q \left(\frac{1}{2}, \hat{\mu} \hat{\lambda} \right) \, , \, 
\end{eqnarray}
for linear or quadratic drag, respectively. Recasting these equations in terms of the original parameters yields:
\begin{eqnarray}
{\cal D}_l(\alpha, \kappa_*) & = & 4 \alpha (1- \alpha) \, {\cal D}_l\left[\frac{1}{2}, 2 (1-\alpha) \kappa_* \right] \, , \\
{\cal D}_q(\alpha, \mu_*) & = & 4 \alpha (1- \alpha) \, {\cal D}_q\left[\frac{1}{2}, 4 \sqrt{\alpha} (1-\alpha)^{3/2} \mu_* \right] , \quad   
\end{eqnarray}
and after substituting (\ref{eq:VGlin}-\ref{eq:VGquad}) on the right-hand side:
\begin{eqnarray}
{\cal D}_l(\alpha, \kappa_*) & = & c_1 \, 4 \alpha (1-\alpha) \exp \left[ \frac{c_2}{2(1-\alpha) \kappa_*} \right] \, , \label{eq:Dl}\\
{\cal D}_q(\alpha, \mu_*) & = & c_3 \, \frac{\alpha^{1/3}}{4^{1/3} (1-\alpha) \mu_*^{4/3}} \, . \label{eq:Dq}
\end{eqnarray}
These expressions provide the dependence of the eddy diffusivity on the relative layer depth $\alpha$ without any additional free parameter as compared to the equal-depth case.
They are valid provided the term ${\cal T}$ can be neglected in equation (\ref{eq:fullttau}). We are now in a position to justify -- in the sense of a scaling estimate -- this assumption by evaluating ${\cal T}$ based on the vortex-gas scaling theory. The vortex-gas scaling theory involves two dominant processes in the temperature equation~\ref{eq:fullttau}: distortion of the background temperature gradient by the meridional barotropic velocity and, crucially, transport of the resulting temperature fluctuations by the barotropic flow, through the term $J\left(\psi, \nabla^2 \tau - \frac{\tau}{\chi^2 \lambda^2}  \right)$.
For a fixed value $\alpha={\cal O}(1)$ of the relative layer depth, the two terms in ${\cal T}$ correspond to the advection of $\nabla^2 \tau$ by the baroclinic velocity $-\bnabla \times (\tau {\bf e}_z)$ and by the base flow velocity $U$, both of which are much smaller than the barotropic velocity $-\bnabla \times (\psi {\bf e}_z)$ because of the strong barotropization of the flow. In other words, the two advective terms in ${\cal T}$ are subdominant as compared to the advective term $J\left(\psi, \nabla^2 \tau - \frac{\tau}{\chi^2 \lambda^2}  \right)$ entering the dominant balance of the vortex-gas scaling theory. Based on this scaling estimate we conclude that ${\cal T}$ indeed appears to be negligible, making the derivation above self-consistent (as confirmed by the numerical results below).


\begin{figure}[h!]
    \centering
    \includegraphics[width=8.6cm]{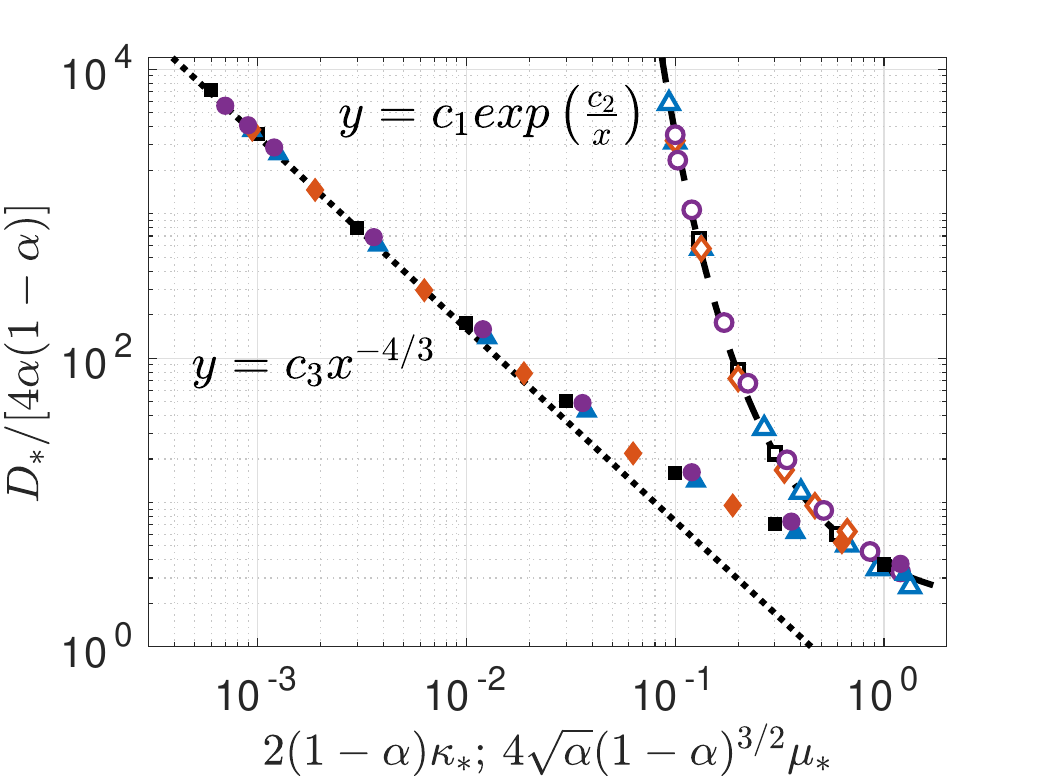}\\
    \caption{Rescaled diffusivity as a function of the rescaled drag coefficient for linear drag (open symbols) and quadratic drag (filled symbols). Same symbols as in figure \ref{fig:raw_data}. The low-drag data collapse onto the vortex-gas predictions (\ref{eq:VGlin}) and (\ref{eq:VGquad}), plotted using $c_1=1.7128$, $c_2=0.7644$ and $c_3=0.3436$.}
    \label{fig:rescaled_D_vs_friction}
\end{figure}

\begin{figure}[h!]
    \centering
    \includegraphics[width=8.6cm]{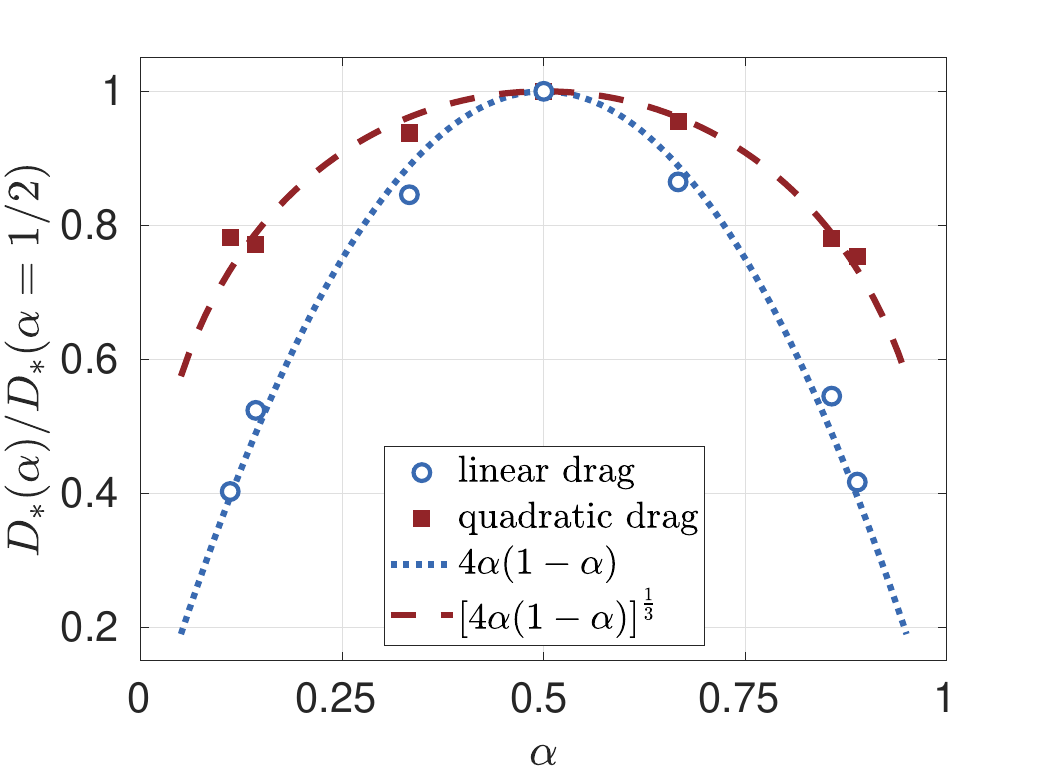}\\
    \caption{Diffusivity coefficient normalized by its equal-depth value for a fixed low value of the drag coefficient acting on the barotropic mode (linear drag, $K=5\times 10^{-2}$; quadratic drag, $M=5\times 10^{-4}$). The lines are the parameter-free predictions (\ref{eq:alpha_dependence_linear}) and (\ref{eq:alpha_dependence_quadratic}).}
    \label{fig:rescaled_D_vs_alpha}
\end{figure}

\textit{Numerical validation.--} 
The theoretical predictions (\ref{eq:Dl}-\ref{eq:Dq}) suggest plotting the rescaled diffusivity $D_*/[4\alpha(1-\alpha)]$ as a function of the rescaled drag coefficient $2(1-\alpha) \kappa_*$ or $4\sqrt{\alpha}(1-\alpha)^{3/2} \mu_*$. As shown in figure \ref{fig:rescaled_D_vs_friction} this representation leads to an excellent collapse of the low-drag numerical data onto two master curves, whose low-drag asymptotic behaviors are captured by the theoretical predictions (\ref{eq:VGlin}-\ref{eq:VGquad}) with excellent accuracy. An alternative way to probe the validity of the predictions (\ref{eq:Dl}-\ref{eq:Dq}) consists in varying the relative layer depth $\alpha$ for fixed drag coefficient acting on the barotropic mode, denoted as $K=(1-\alpha)\kappa_*$ for linear drag and $M=(1-\alpha)\mu_*$ for quadratic drag. Specifically, in Figure \ref{fig:rescaled_D_vs_alpha} we plot $D_*$ normalized by its value for the equal-depth case $\alpha=1/2$ and the same value of $K$ or $M$. Based on (\ref{eq:Dl}-\ref{eq:Dq}) we expect this ratio to be given by the parameter-free predictions: 
\begin{eqnarray}
 & & \frac{{\cal D}_l(\alpha, K/[(1-\alpha)])}{{\cal D}_l(1/2, 2K) } = 4 \alpha (1-\alpha)  \, ,\label{eq:alpha_dependence_linear}\\
 & & \frac{{\cal D}_q(\alpha, M/[(1-\alpha)]) }{{\cal D}_q(1/2, 2M) } = [4 \alpha (1-\alpha)]^{1/3}  \, ,
 \label{eq:alpha_dependence_quadratic}
\end{eqnarray}
for linear and quadratic drag, respectively. The data in Figure \ref{fig:rescaled_D_vs_alpha} are in excellent agreement with these predictions for both types of drag.

\ \\

\textit{Conclusion.--}We have determined the dependence of the eddy diffusivity induced by baroclinic turbulence in the two-layer model with arbitrary layer depths. The derivation hinges on the barotropic-baroclinic decomposition, which we combined with the vortex-gas scaling theory. 
The predictions include no additional free parameters as compared to the equal-depth case. They are validated quantitatively by the numerical simulations, which indicates the robustness of the vortex-gas theory. In the context of the parameterization of transport by ocean mesoscale turbulence, a theory for the unequal-depth case allows one to properly account for the strongly non-uniform ocean stratification. This brings us one step closer to being able to project fully three-dimensional patches of ocean~\cite{meunier2023direct,meunier2023vertical} onto their first two vertical modes, with the goal of inferring the overall transport properties of an ocean patch based on the associated two-layer description.\newline

\textit{Acknowledgments.--}This research is supported by the European Research Council under Grant Agreement FLAVE 757239. We thank Julie Meunier for insightful comments.


%

\end{document}